\documentclass[namedreferences]{solarphysics}
\usepackage[optionalrh,solaenum]{spr-sola-addons} 
\usepackage{graphicx}                    
\usepackage{color}                       
\usepackage{url}                         

\newcommand{\aap}{    {\it Astron. Astrophys.}}

\newcommand{\apj}{    {\it Astrophys. J.}}

\newcommand{\grl}{    {\it Geophys. Res. Lett.}}

\newcommand{\jgr}{    {\it J. Geophys. Res.}}

\newcommand{\nat}{    {\it Nature}}

\newcommand{\solphys}{{\it Solar Phys.}}

\newcommand{\ssr}{    {\it Space Sci. Rev.}}

\begin{document}

\begin{article}

\begin{opening}

\title{Eclipses observed by LYRA - a sensitive tool to test the models for the solar irradiance}

%
\author{A.I.~\surname{Shapiro}$^{1}$\sep
              W.~\surname{Schmutz}$^{1}$\sep
              M.~\surname{Dominique}$^{2}$\sep 
              A.V.~\surname{Shapiro}$^{1,3}$     
       }

\runningauthor{Shapiro et al.}
\runningtitle{Eclipses observed by LYRA}

%
  \institute{$^{1}$ Physikalisch-Meteorologishes Observatorium Davos, World Radiation Center, 7260 Davos Dorf, Switzerland\\
                     email: \url{alexander.shapiro@pmodwrc.ch}\\ email: \url{werner.schmutz@pmodwrc.ch} \\  email: \url{anna.shapiro@pmodwrc.ch}  \\
            $^{2}$ Royal Observatory of Belgium, Ringlaan 3 B-1180 Brussel, Belgium\\
                     email: \url{Marie.Dominique@oma.be} \\
             }

\begin{abstract} 
We analyze the light curves of the recent solar eclipses measured by the Herzberg channel (200--220 nm) of the Large Yield RAdiometer (LYRA) onboard PROBA-2. The measurements allow us to accurately retrieve the center-to-limb variations  { (CLV)} of the solar brightness. The formation height of the radiation depends on the observing angle so the examination of the CLV provide information about a broad range of heights in  the solar atmosphere.   We employ the 1D NLTE radiative transfer COde for Solar Irradiance (COSI) to model the measured light curves and corresponding CLV dependencies.  The modeling is used to test and constrain the existing 1D models of the solar atmosphere, e.g. the temperature structure of the photosphere and the treatment of the pseudo-continuum opacities in the Herzberg continuum range.   We show that COSI can accurately reproduce not only the irradiance from the entire solar disk, but also the measured CLV.  It hence can be used as a reliable tool for modeling the variability of the spectral solar irradiance.
\end{abstract}

%

\end{opening}

\section{Introduction.}\label{sect:intro} 
The variability of the solar irradiance may have a direct impact on climate (see e.g. the recent reviews by \opencite{haigh2007} and \opencite{grayetal2010}). Although the measurements and modeling of the solar irradiance were under the close attention during the last decade, the complete picture of the solar variability is still far from being clear (see e.g. \opencite{harderetal2009}; \opencite{haighetal2010}). Therefore the launch of every new space mission devoted to the measurements of the solar irradiance is able to provide a crucial piece of  complementary information as well as to nourish the theoretical models.

\begin{figure} 
\centerline{\includegraphics[width=1.0\textwidth,clip=]{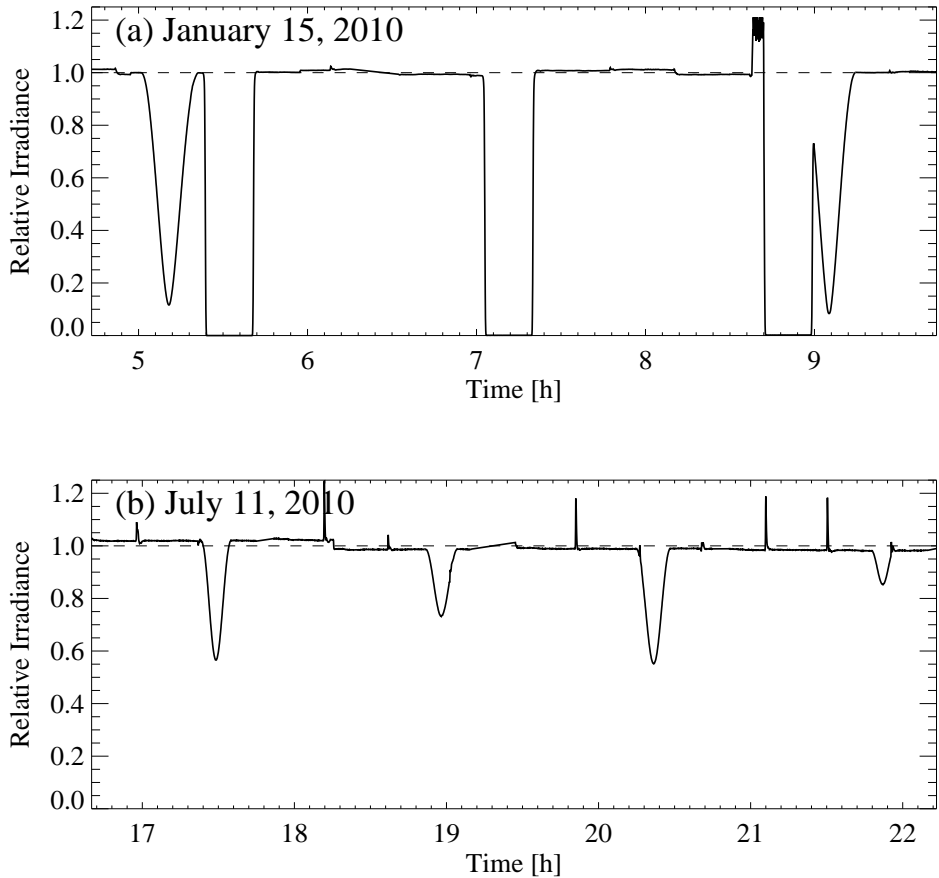}}
\caption{Relative variations of the irradiance as measured by the Herzberg channel of LYRA. The intervals of zero intensity occur during the occultations when PROBA-2 passes the Earth  shadow in the winter season. The panels show two eclipses separated by the three occultations  on January 15, 2010 (a) and four eclipses on July 11, 2010 (b). The periodic abrupt changes of the irradiance level are due to the spacecraft maneuvers. }
\label{fig:intr}
\end{figure}

In this paper we analyze the first measurements of the Large Yield RAdiometer (LYRA) \cite{LYRA1,LYRA2}  onboard the PROBA-2 satellite launched on November 2, 2009. Up to now LYRA has observed several solar eclipses (see Fig.~\ref{fig:intr}). 

 During the eclipse the Moon consecutively covers different parts of the solar disk. The light curve of the eclipse depends on the CLV of the solar brightness and on the geometry of the eclipse (the angular radii of the Sun and the Moon as well as the minimum distance between their centers which is reached during the maximum phase of the eclipse). If the geometry of the eclipse is known and the distribution of solar brightness has radial symmetry then the light curve of the eclipse can be used to retrieve the CLV of the solar brightness. Let us notice that the assumption of the radial symmetry is well-justified for the January 15, 2010 eclipse as the solar activity level was very low (according to the USAF/NOAA data the total sunspot area was about 0.025\% of the full solar disk). The CLV of the solar brightness  provide a valuable information about the solar atmosphere (\opencite{prietoetal2004};  \opencite{koesterkeetal2008}) and determine the irradiance variations on the time-scale of the solar rotation \cite{fliggeetal2000}. Additionally the changes of the spectral solar irradiance during the eclipses are important for studying the Earth's atmosphere response  (\opencite{ecl_climate1}; \opencite{ecl_climate2}).

\opencite{shapiroetal2010} showed that the 1D NLTE radiative transfer COde for Solar Irradiance (COSI) (see \opencite{haberreiter2008}) allows to calculate the solar spectrum (125 nm--1$\,{\rm \mu} $m) from the entire solar disc with a high accuracy. In this paper we use COSI to calculate the CLV of the solar brightness and compare them with ones deduced from the eclipse light curves as observed by LYRA. We show that the measured CLV provide an important constrains on the UV opacities and the temperature structure of the solar atmosphere.   We come up with a model which allows to accurately reproduce the measurements.

We restrict ourselves to the modeling of the eclipse profiles and CLV in the Herzberg channel of LYRA. The solar irradiance  in the Herzberg continuum range (200--220 nm)  is of especial importance for the climate modeling as it directly affects the ozone concentration and stratospheric temperature \cite{brasseuretal1997,rozanovetal2006,anna2011}. The proper modeling of the formation of this radiation in the solar atmosphere is a base for the variability modeling and for the irradiance reconstruction to the past (see e.g. \opencite{krivovaetal2011}; \opencite{vieiraetal2011}; \opencite{SSIrec}).

We are aware that the 1D models do not necessarily reflect the average physical properties of the inhomogeneous solar atmosphere \cite{uitenbroek2011}. Thus the main goal of this paper is  not to learn the new facts about the dynamic 3D solar atmosphere  but rather to develop a reliable semi-empirical tool for modeling the solar irradiance variability.

In Sect.~\ref{sect:empirical} we deduce the empirical CLV from the LYRA observations of the January 15, 2010 eclipse. In Sect.~\ref{sect:modeling} we compare these CLV with ones calculated by COSI and discuss the constrains on the temperature structure of the solar atmosphere (Sect.~\ref{subsect:T}) and UV opacities (Sect.~\ref{subsect:FALC}). The main results  are summarized in Sect.~\ref{sect:conc}.

\section{Empirical center-to-limb variations as deduced from the LYRA observations}\label{sect:empirical}
 { PROBA-2 evolves on a dawn-dusk heliosynchronous orbit, with an altitude of 720 km on average, which allows a quasi-permanent observation of the Sun. The spacecraft accomplishes a full orbit in about 100 minutes. In case of solar eclipse, it is therefore not unusual that it crosses the eclipse zone more than once.
Lyra data in Herzberg channel are acquired, for all three units, by experimental PIN detectors made of diamond. Such detectors have proven to be very stable with respect to temperature variations \cite{LYRA_detectors}.  The nominal cadence of acquisition is 20 Hz. A more detailed discussion of the in-flight performance of LYRA is given in \opencite{marie2011}.}

\begin{figure} 
\centerline{\includegraphics[width=1.0\textwidth,clip=]{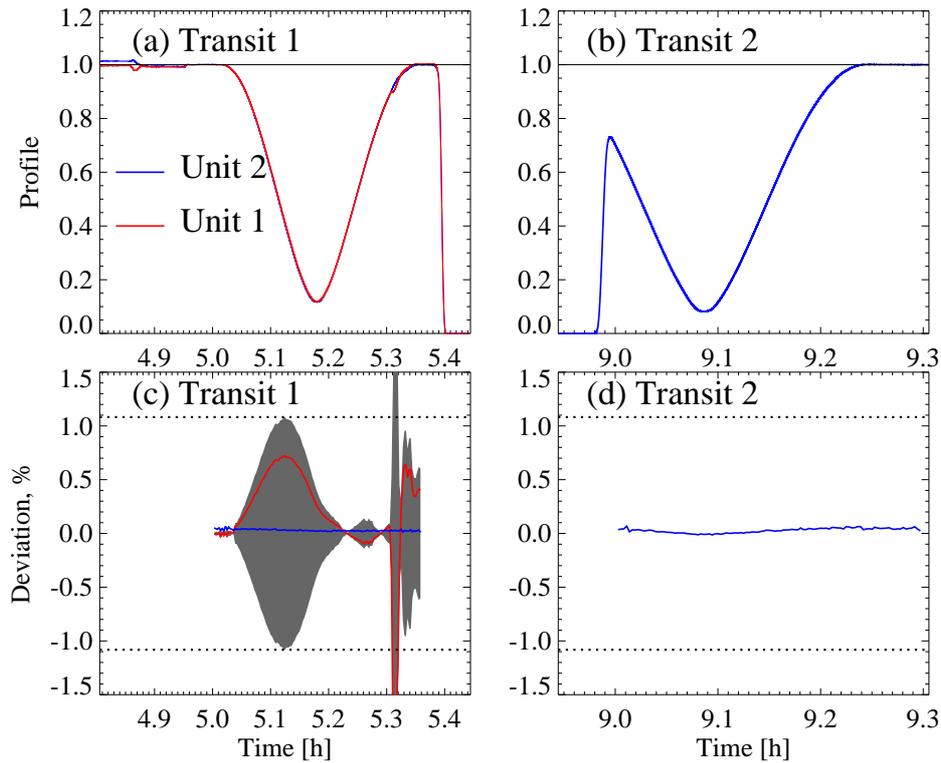}}
\caption{The profiles of the 05:00 (a) and 09:00  (b) transits  retrieved from the level 1 data as well as the deviations between these profiles and the profile retrieved from the level 2 data of the unit 2 (c,d). 
 { The shaded area on panel c indicates the estimated error range.}
The  dotted lines correspond to the widest part of the error range in the time interval  { between 5 h and  5.3 h (i.e. excluding the problematic feature after 5.3 h).}}
\label{fig:ecl1_obs}
\end{figure}

The first light curve of the eclipse event was obtained by LYRA on January 15, 2010.
The eclipse was shortly preceded by  LYRA first light on January 6, 2010 and was the longest annular solar eclipse of the millennium. 
It was observed on the ground from Africa and Asia and was seen as a partial from the PROBA-2. The eclipse lasted more than 6 hours,  { so the PROBA-2 passed through the Moon's shade three times.} However the intermediate transit could not be observed due to the simultaneous occultation (i.e. it was shaded by the Earth). 

The raw (level 1) data collected by the Herzberg channel of LYRA during this eclipse are presented in panels a and b of Fig.~\ref{fig:ecl1_obs}. The plotted data were corrected for the dark current which is still present in the original data. The 05:00  UTC transit  { of the January 15, 2010 eclipse}  (hereafter first transit) was simultaneously observed by the LYRA units 1 and 2 (the back-up and standard acquisition units, accordingly), while the 09:00 UTC transit (hereafter second transit) only by the unit 2. The drops of the irradiance after the first and before the second transits correspond to the occultations (see panel a of Fig.~\ref{fig:intr}).

 { The level 1 data are uncalibrated. The calibrated (level 2) data are also available for the community. For the analysis presented below,  we will therefore use the profiles retrieved from the level 2 data.   These data are corrected for the temperature effects, degradation and the dark current. Let us note that level 2 data always refer to the unit 2 measurements, while the measurements from the back-up units 1 and 3 are normally  used  for the calibration and currently only available  in their uncalibrated version \cite{marie2011}. Originally the level 2 data were corrected for the degradation by adding a time-dependent offset (to allow a better analysis of the solar flares). The offset shifts the zero-level of the irradiance and leads to the erroneous profiles of the eclipses and solar variability \cite{anna_eclipse}. Therefore for our analysis the offset was removed from the level 2 data. }

 { The blue curves on the panels c and d of Fig.~\ref{fig:ecl1_obs}  represent} the deviations between the level 1 and level 2 data. The zero-level of both datasets was corrected as discussed above. One can see that the level 2 data are almost identical to the original measurements of the standard acquisition unit 2.  {  This confirms that the temperature correction for the diamond detectors is very small. From now on we will not distinguish between the level 1 and level 2 data of the unit 2. On the contrary there is a significant deviation between the data  collected by the unit 1 and unit 2 during the first transit.  Both units were carefully calibrated in the ground facilities so their different responses are connected with the degradation which affects the sensitivities of the units. The standard acquisition unit 2 was opened almost constantly, while the back-up units 1 and 3 were opened only occasionally. As a result on January 15, 2010 unit 2 degraded approximately by  15\% while unit 1 did not yet show any noticeable degradation. The differential behavior of the units allows us  to estimate the error range of the measured profiles. We defined the uncertainty that it is twice the difference between the profiles as observed by the unit 2 and the unit 1 (see panels c of Fig.~\ref{fig:ecl1_obs}). This approach does not include any systematic deviations, common to both units but it measures the degradation due to the space exposure.  
The arbitrary character of this estimate does  not affect the results presented below.} The second transit was only observed by the unit 2 so for its analysis we will use the error range defined from the first transit.

The profile of the irradiance variation during the eclipse depends on the CLV of the solar brightness. For example a strong decrease of the brightness towards the solar limb would lead to a smaller residual irradiance during the maximal phase of the eclipse and accordingly to a deeper eclipse profile. Thus, the observed profiles allow to determine the CLV. For simplicity we adopted the widely used polynomial parametrization of the CLV (see e.g. \opencite{neckel1994}; \opencite{neckel2005}):
\begin{equation}
\frac{I  (\mu)}{I_{\rm center}}=\sum\limits_i A_i \mu^i,
\label{eq:param}
\end{equation}
where $I_{\rm center}$ is the disk-center intensity, $\mu$ is the cosine of the heliocentric angle and $A_i$ are free parameters. The solar disk was divided into the thirteen supposedly uniform concentric rings and the brightness of each ring was calculated with the help of the Eq.~(\ref{eq:param}) using the $\mu$ value of the ring's mean circle. We checked that such division of the solar disc allows to calculate the irradiance profiles with the accuracy better than $0.02 \%$ and thus is sufficient for our purposes.  Simple geometrical calculations allow to obtain the profile of the eclipse for each set  of the coefficients, assuming that the time dependencies of the angular distance between the Sun and the Moon as well as of their angular sizes  are known. 

For the both transits of the January 15, 2010 eclipse we searched for the set of the  coefficients  $A_i$ which lead to the best agreement with the observed profiles and minimized the  error ${\cal E}$ defined as:
$${\cal E} = \sum\limits_i {\left (  {\cal P}_{\rm obs}^i -  {\cal P}_{\rm emp}^i\right )^2 }, $$
where $ {\cal P}_{\rm obs}$ is the observed profile and  ${\cal P}_{\rm emp}$ is the empirical profile calculated from the Eq.~(\ref{eq:param}).
For the first and second transits the summation was  done over the time intervals between   { 5 h and 5.3 h and  between 9 h and  9.25 h respectively}.

 { The minimization was performed applying the additional condition of the monotonous CLV.}  The CLV profiles were calculated using the second degree polynomial. The employment of the higher degree polynomial (up to sixth degree) had no visible effect. The CLV depend on the wavelength \cite{neckel1996,hestrofferandmagnan1998}. Thus, the coefficients  $A_i$  determined by the minimization procedure correspond to the CLV of the solar intensity convolved with the profile of the LYRA Herzberg channel. The latter is a combined profile of the detector and filter.

\begin{figure} 
\centerline{\includegraphics[width=1.0\textwidth,clip=]{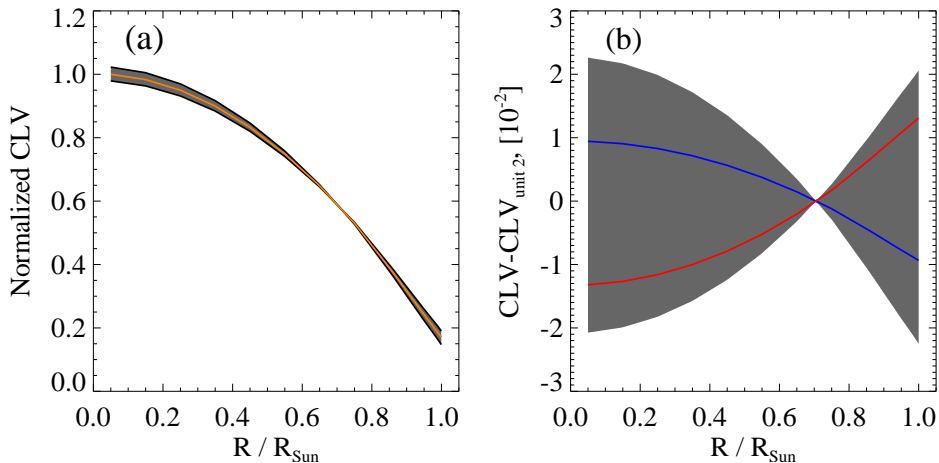}}
\caption{ { (a) Empirical CLV deduced from the profile of the first transit as observed by the unit 2 (orange curve); (b) the deviations between the CLV plotted on panel (a) and  the CLV deduced from the profile of the first transit as observed by the unit 1 (red curve) and CLV deduced from the profile of the second transit as observed by the unit 2 (blue curve). The empirical CLV deduced from the two extreme profiles corresponding to the edges of the error  range in Fig.~\ref{fig:ecl1_obs}  delimit the CLV error range (the shaded area).}}
\label{fig:ecl1_emp_CLV}
\end{figure}

\begin{figure} 
\centerline{\includegraphics[width=1.0\textwidth,clip=]{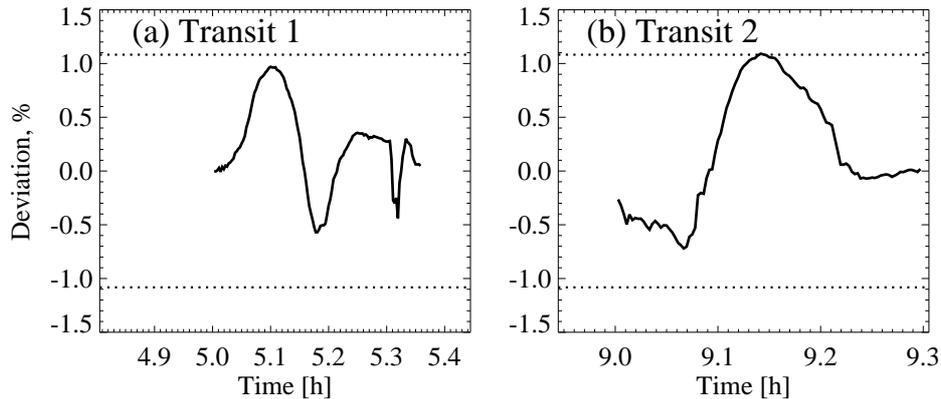}}
\caption{ { The deviations between the profiles as measured by unit 2 and as calculated with the empirical CLV. The error bars (dotted lines) are the same as in  Fig.~\ref{fig:ecl1_obs}. }}
\label{fig:ecl1_emp}
\end{figure}

 { The resulting empirical CLV dependencies are presented in  Fig.~\ref{fig:ecl1_emp_CLV}.  The CLV dependency for the first transit} is normalized to unity in the disk center, while all other CLV dependencies are normalized to give the same integral flux from the entire disk.
The error range of the CLV was estimated performing the minimization procedure  to the two  maximal error profiles of the first transit. So the shaded area in the Fig.~\ref{fig:ecl1_emp_CLV} is constrained by the two CLV dependencies which correspond to the edges of the  shaded area in the panel c of Fig.~\ref{fig:ecl1_obs}. 

 { The deviations between the profiles as measured by LYRA  and calculated with the empirical CLV  are shown in Fig.~\ref{fig:ecl1_emp}. The deviations can be attributed to the limited accuracy of the measurements (note that they are within the estimated error bars) and to the violation of the radial symmetry of the solar brightness. One of the sources of the asymmetry of the solar brightness  is the inhomegenous structure of the quiet Sun, which consists from several  brightness components \cite{fontenlaetal1999}. The small amplitude of the deviations supports the analysis presented below.}

\section{Comparison with modeling}\label{sect:modeling}
To calculate the theoretical CLV we employed the 1D NLTE radiative transfer code COSI developed by \opencite{hubeny1981}; \opencite{hamannschmutz1987};  \linebreak \opencite{schmutzetal1989}; \opencite{haberreiter2008}; \opencite{shapiroetal2010}. COSI simultaneously solves the statistical equilibrium and radiative transfer equations in the spherically symmetrical geometry. The temperature and density structures of the different components of the solar atmosphere were taken from \opencite{fontenlaetal1999}, while the electron density and all level populations were self-consistently calculated in the NLTE.

\subsection{Solar irradiance in the Herzberg continuum range}\label{subsect:Herzberg}
The proper calculation of the radiative transfer in the Herzberg continuum region is sophisticated by two factors.
Firstly, the continuum opacity in this region is strongly affected by the NLTE overionization (see e.g. \opencite{ShchukinaandTrujillo2001}; \opencite{shorthauschildt2005}), so the proper NLTE calculations are necessary. 
Secondly, the immense number of weak, mostly spectrally unresolved lines, form the so-called UV line haze in this region. The creation of the correct line list which would include all possible spectral lines is a task of a tremendous difficulty. Although considerable progress was reached during the last few decades, there is still only 1\% of the UV lines which are measured in the laboratory, while all remaining lines are predicted only theoretically \cite{kurucz2005}. As the consequence the existing line list are not complete and underestimate the opacity in the UV
spectral region \cite{busaetal2001,haberreiter2008,shorthauschildt2009}. 

To account for the missing opacity \opencite{shapiroetal2010} multiplied the continuum opacity coefficient  $ k_c (\lambda)$ by the  wavelength dependent coefficient $f_c(\lambda)$:
\begin{equation}
k_c' (\lambda)= k_c (\lambda) \cdot f_c (\lambda) .
\label{eq:factor}
\end{equation}
The  employed multiplicative coefficient is a step  function of the wavelength with the step equals to 1 nm.  It  was empirically determined so that the UV irradiance calculated by COSI equals to the irradiance as measured by SOLSTICE (SOLar-STellar Irradiance Comparison Experiment,  see \opencite{SOLSTICE}) onboard the SORCE satellite \cite{rottman2005} during the 2008 solar minimum.  
The additional opacity which is necessary to reproduce the SOLSTICE/SORCE irradiance does not exceed a few percents of the total opacity included in COSI. It is only necessary in the 160--320 nm spectral region.  Shortward of 160 nm the total opacity is mainly dominated by the photo-ionization continuum opacity, while longward of 320 nm the existing line lists are accurate enough to reproduce the irradiance with better than 5\% accuracy (and with better than 2\% accuracy longward of 400 nm). 

As during the solar minimum the solar irradiance (except the extreme UV) is dominated by the quiet Sun,  the solar atmosphere model C (average supergranule cell interior model) from \opencite{fontenlaetal1999} was used for the calculations. Let us however note that it would be possible to be in agreement with the SOLSTICE/SORCE measurements while using another temperature profile by adjusting additional opacities.  Thus the SOLSTICE/SORCE measurements  alone does not allow to constrain the temperature structure of the solar atmosphere. In Sect.~\ref{subsect:T} we show that although simultaneous adjustment of the temperature structure and additional opacities does not affect the irradiance from the entire solar disc, it significantly changes the CLV. The comparison with the measured CLV allows to choose the most suitable for the CLV calculations  model.

 \opencite{shapiroetal2010} assumed that the coefficient of the additional opacity does not depend on height in the solar atmosphere.  This implies that it does not have  any dominant source  as the concentration of every particular ion or molecule is height dependent. In Sect.~\ref{subsect:FALC} we show that the CLV provide important complementary information which allow to reevaluate this assumption and help to understand the nature of the additional opacity.

\subsection{Test of the temperature structure}\label{subsect:T}
To test the sensitivity of the CLV to the change of the temperature structure we performed the calculations employing the models for several different components of the solar atmosphere: model A (faint supergranule cell interior), model C (average supergranule cell interior), model E (quiet network) and model P (plage). The temperature and density structures were taken from \opencite{fontenlaetal1999}. The models A and E correspond to the cold and warm components of the quiet Sun, while the Model C represents the spatially averaged quiet Sun. \opencite{shapiroetal2010} showed that the calculations with the latter model can reproduce spectral irradiance measured by SOLSTICE (up to 320 nm) and SIM (Solar Irradiance Monitor; \opencite{SIM})  (from 320 nm onward) onboard the SORCE satellite during the 2008 solar minimum, as well as SOLSPEC (SOLar SPECtral Irradiance Measurements; \opencite{thuilleretal2004}) during the ATLAS 3 mission in 1994 with high accuracy. They used the model  C to calculate the $f_c(\lambda)$ factor (see Eq.~(\ref{eq:factor})) for the additional opacities in the 160--320 nm spectral range. 

\begin{figure} 
\centerline{\includegraphics[width=1.0\textwidth,clip=]{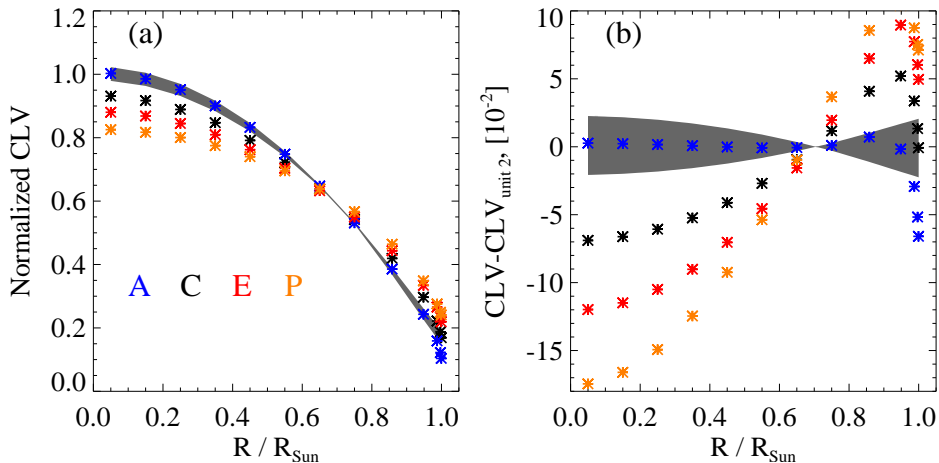}}
\caption{The same as Fig.~\ref{fig:ecl1_emp_CLV}  but with the CLV dependencies calculated for  the faint supergranule cell interior(A), average supergranule cell interior (C), quiet network (E) and plage (P).}\label{fig:CLVmodels}
\end{figure}

The same procedure of the $f_c(\lambda)$ factor fitting was performed for the Models A and E. The Model A is colder than Model C so it yields smaller UV irradiance. Thus a smaller additional opacity is necessary to reproduce the SOLSTICE/SORCE measurements in the 160-320 nm spectral region. Accordingly the use of  Model E leads to a larger additional opacity. The Model P yields so high irradiance that it is not possible to reach SOLSTICE/SORCE level by increasing the continuum opacity. Thus  we didn't recalculate  the $f_c(\lambda)$  factor for the model P and left it the same as for the Model C. 

In Fig.~\ref{fig:CLVmodels} we present the calculated CLV for each of these models. The radiation which comes from the regions close to the solar limb is formed in higher and colder regions of the solar atmosphere than the radiation coming from the disc center. Hence the solar brightness is decreasing towards the limb. One can see that the colder the model, the stronger CLV it yields. This can be partly explained by the fact that the sensitivity of the Planck function to the temperature change is the decreasing function of the temperature.  Thus the same change of  temperature results in larger alteration of the Planck function and accordingly larger change of the emergent irradiance for the colder models.
The complete picture depends  on the temperature and density structure as well as on the opacity behavior in each of the solar atmosphere components. It is further sophisticated by the NLTE effects which causes deviations of the source function from the  Planck function.

Although due to the readjustment of the additional opacity the calculations with the Models A and E yield the same UV irradiance as the calculation with the Model C, the corresponding CLV dependencies are remarkably different. All differences between the theoretical and empirical CLV's have a sudden drop at $R/R_{\rm Sun} \approx 0.99$ (see panel (b) of Fig.~\ref{fig:CLVmodels}).  The rings which correspond to these points  (see Sect.~\ref{sect:empirical}) have very small relative area, so the eclipse profiles are basically insensitive to the change of their brightness.  Thus the reliability of the empirical CLV for these points is very low.

Interestingly, the calculations with the model C underestimate the CLV and are outside of the estimated error region. One of the possible reasons for this could be the erroneous assumption of the depth independency of the additional opacity coefficient $f_c$ (see Eq.~(\ref{eq:factor})). This possibility will be discussed in the Sect.~\ref{subsect:FALC}. On the other hand it is known that
  the 1D models can underestimate the anisotropy of the radiation field \cite{CN_NLTE,kleintetal2011} and the CLV of the solar brightness \cite{koesterkeetal2008,uitenbroek2011}.   Hence the reported disagreement could be a signature of the general  problems which are inherent to 1D modeling. 

The warmer Models E and P  yield even weaker CLV than the Model C. At the same time the colder model A yields the CLV which are in a good agreement with the empirical ones.

 \begin{table}
 \caption{Theoretical profiles  vs. observed and   { empirical  (calculated with the empirical CLV)} profiles. The numbers are discrepancies calculated with the help of Eq.~(\ref{eq:delta}). The minimum values  of the discrepancy are boldfaced.}\label{table:comp}
 \begin{tabular}{c  c  c  c  c}     
\hline
     &  \multicolumn{2}{c}{Passage 1} &  \multicolumn{2}{c}{Passage 2} \\
     &  observed &   { empirical}  & observed &  { empirical}\\
 A &   { 49.9}              &      { 3.3}   	    & 	 	 { 65.0}   &   { 9.9}    \\
 C &  105.8            &     92.5            & 		130.1		&  113.0    \\
 E &  168.7            &    160.4 	   & 		197.3	&  186.5  \\
 P &  216.3            &    209.2 	   & 		245.2	&  236.3  \\
 \hline
\end{tabular}
\end{table}
\begin{figure} 
\centerline{\includegraphics[width=1.0\textwidth,clip=]{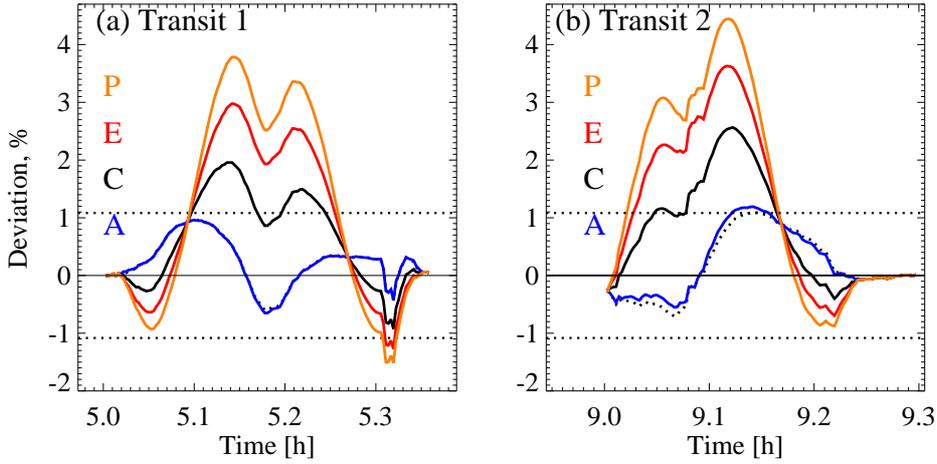}}
\caption{The same as Fig.~\ref{fig:ecl1_emp} but with the eclipse's profiles calculated for the solar atmosphere  models A, C, E and P. }\label{fig:Pmodels}
\end{figure}

The corresponding deviations between the calculated and measured profiles are shown in Fig.~\ref{fig:Pmodels}.  { The structure common to the both transits represents properties of the models, while individual structure is due to the limited accuracy of the measurements and   violation of the radial symmetry of the solar brightness. } We can define the  discrepancy between the profiles ${\cal P}_1^i$ and ${\cal P}_2^i$  as: 
\begin{equation}
\Delta_{1,2}=10^{-4} \, \sqrt{  \left (   \sum\limits_{i=1}^{N} \left ( {\cal P}_1^i-{\cal P}_2^i \right )^2/N  \right )},
\label{eq:delta}
\end{equation}
where N is number of the time points in the profiles and the scaling factor $10^{-4}$ is introduced for convenience. 

In Table~\ref{table:comp} we present the discrepancies between  { the theoretical, measured and empirical (calculated with the empirical CLV, see Sect.~\ref{sect:empirical})  profiles}. One can see that the calculations with the Model A are in the good agreement with the measurements and is very close to the  { empirical}  profile. Let us however note that the model A is not able to properly reproduce the visible and near infrared irradiance as well as the main molecular bands (e.g. CH G-band and CN violet system) in the solar spectrum. The calculations with the Models E and P result in too weak CLV and consequently in too high residual irradiance during the maximal phase of the eclipse.  Although the calculations with the Model C are outside of the estimated error range the yielded profiles are reasonably close to the observations (e.g. they will be hardly distinguishable from the measured ones in the scale of the Fig.~\ref{fig:intr}). Thus  the Model C still can be used in the calculations when the high accuracy is not necessary.

\subsection{Adjustments of the additional opacities}\label{subsect:FALC}
The introduction of the additional opacity described in Sect.~\ref{subsect:Herzberg} changes the formation heights of the UV radiation and thus affects the CLV. All calculations presented in Sect.~\ref{subsect:T} were performed assuming depth independent  coefficient $f_c$ in Eq.~(\ref{eq:factor}). At the same time if the additional opacity arises from the unaccounted lines of some particular  atom or molecule X then it should be scaled with the relative concentration of X.   Then the  coefficient $f_c$ becomes height dependent:
 \begin{equation}
f_c(\lambda, h)=1+{\cal F}_c  (\lambda) \cdot \frac{n_X (h) / n_{\rm total} (h) }   {  \; \; \;  \; \; {      \left (      n_X  (h)/ n_{\rm total} (h)       \right )             }_{\rm max}               },
\label{eq:factorH}
\end{equation}
where $h$ is the height in the solar atmosphere,  $n_{\rm total}$ and $n_X$ are the total and species X  concentrations respectively. ${\cal F}_c  (\lambda)$ is a step function which can be determined empirically in the same way as the $f_c$ factor was determined in \opencite{shapiroetal2010}. The height dependency of the additional  opacity is introduced by the fraction in the right side of the equation. The latter is  the relative concentration of the species X, normalized   for convenience to unity.

\begin{figure} 
\centerline{\includegraphics[width=1.0\textwidth,clip=]{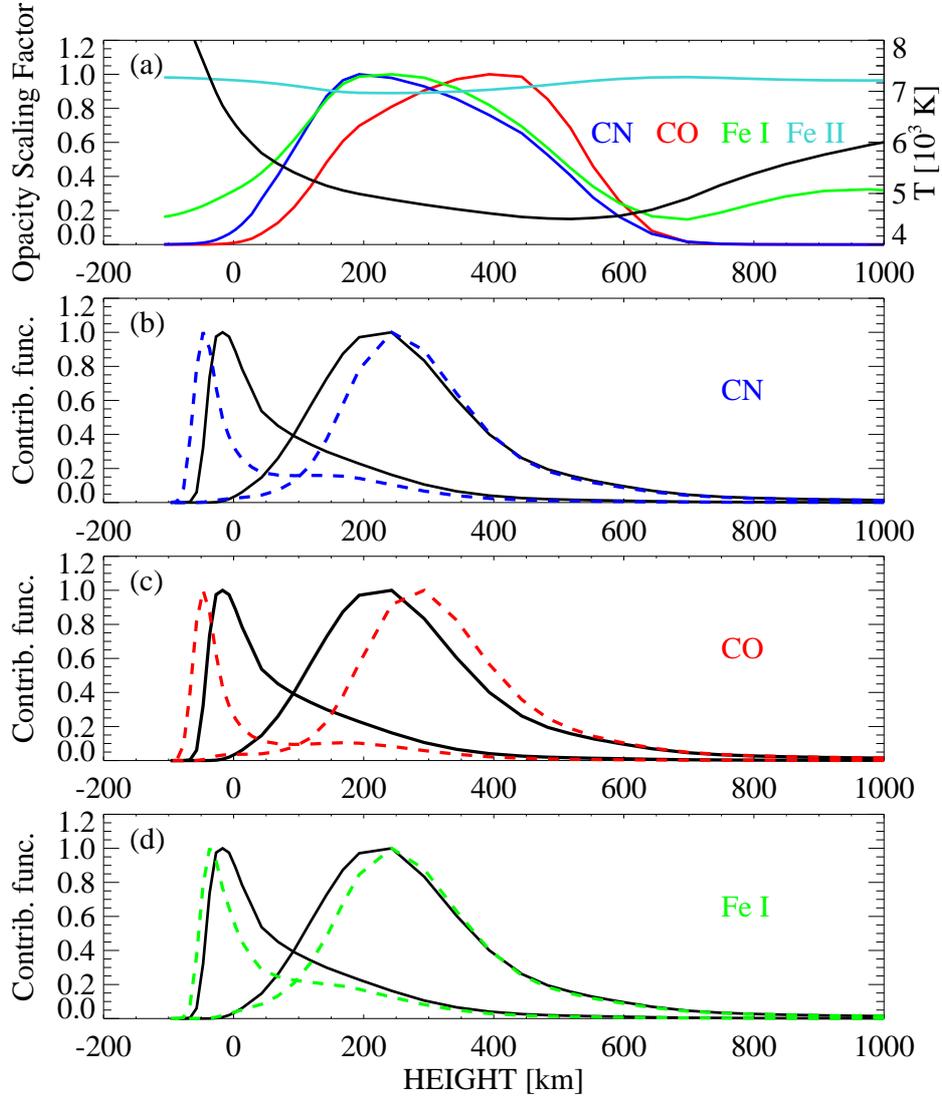}}
\caption{Opacity scaling factor (the fraction in the right side of  the Eq.~(\ref{eq:factorH})) and temperature (black curve) as a function of height  (a) as well as the contribution functions for  near the limb ($R/R_{\rm Sun}=0.95$)  and disc center intensities. The contribution functions are shown for the case of height independent additional opacity (black curves in (b), (c) and (d)) and additional opacity  scaled with the CN (b), CO (c) and Fe I (d) relative concentrations. The zero-point of the height scale  is defined as the layer at which the continuum optical depth at 500 nm is equal to one.   }\label{fig:contr}
\end{figure}

To better understand the possible origin of the additional opacity we scaled it with the relative concentrations of CN and CO molecules and Fe I ion. In each case we empirically determined the ${\cal F}_c  (\lambda)$ factor to reproduce the SOLSTICE/SORCE UV measurements. In panel a of Fig.~\ref{fig:contr} we show the dependency of the fraction from the  Eq.~(\ref{eq:factorH}) on height. The CN and CO molecules are mainly present in the narrow photospheric layer. As the CO molecule has larger dissociation potential than the CN (11.1 eV against 7.76 eV) it is more sensitive to the temperature change \cite{berdyuginaetal2003} and the maximum of the CO relative concentration is very close to the temperature minimum. The maximum of the CN relative concentration is slightly shifted towards the lower levels of the photosphere where the effect of the density increase overcompensate the temperature increase. Iron is mostly ionized throughout the solar atmosphere. Similarly to the molecular case the neutral iron concentration has a peak slightly below  the temperature minimum. At the same time it does not drop so abruptly in the lower level of the solar photosphere and starts to increase in the chromosphere due to the strong density decrease. 

On the three lower panels of Fig.~\ref{fig:contr}  we show the contribution functions for the intensities in the Herzberg continuum range. They were obtained by convolving the individual contribution functions  
(see, e.g. \opencite{gray1992}, p.\,151) for every frequency with the combined profile of the detector and filter in the Herzberg channel. Thus the plotted contribution functions  sample the regions where the  irradiance as measured by the Herzberg channel of LYRA is formed. The scaling of the additional opacity with the Fe I or molecular concentrations alters the optical depth scale and simultaneously shifts the limb and disc center contribution functions to the higher and lower levels of the solar atmosphere respectively.  This increases the distance between the peaks of the limb and disc center contribution functions, which results in the stronger CLV. The effect is the most prominent for the scaling with the CO relative concentration (due to the strong shift of the limb contribution function) and the least prominent for the Fe I case (due to a small effect on the disc center contribution function).

\begin{figure} 
\centerline{\includegraphics[width=1.0\textwidth,clip=]{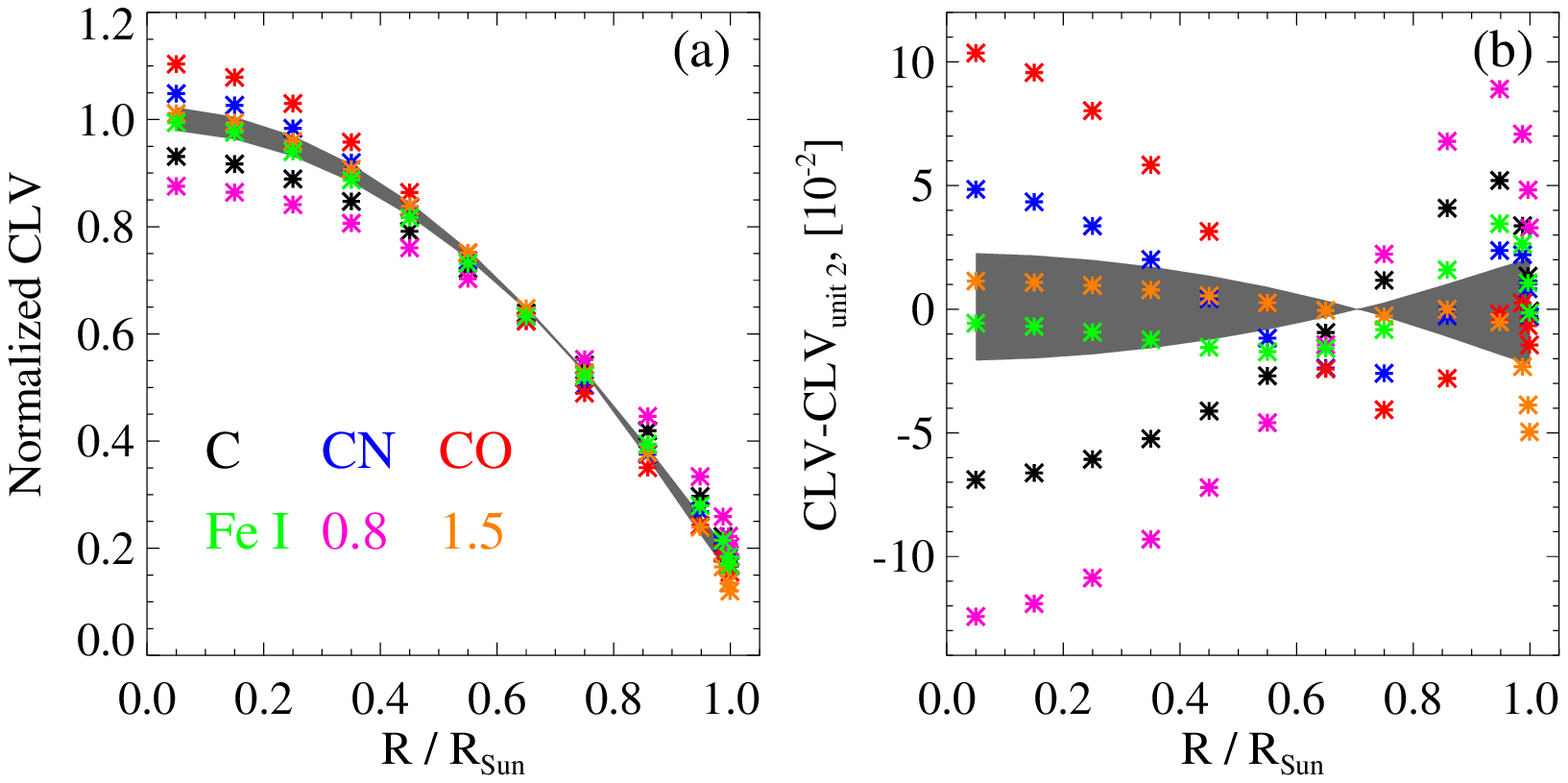}}
\caption{The same as Fig.~\ref{fig:ecl1_emp_CLV} but for different modes of calculations with the Model C. The ``CN'',  ``CO'' and ``Fe I'' points were calculated scaling the additional opacity with the relative concentration of the corresponding species.  The ``0.8'' and ``1.5'' points were calculated adjusting the  additional opacity so that COSI yields 80\% and 150\% of the  UV irradiance measured by the SOLSTICE/SORCE. The ``C'' points correspond to the standard calculations using Model C (i.e. with height independent additional opacities adjusted to reproduce the SOLSTICE/SORCE UV measurements).   }\label{fig:FALC}
\end{figure}

\begin{figure} 
\centerline{\includegraphics[width=1.0\textwidth,clip=]{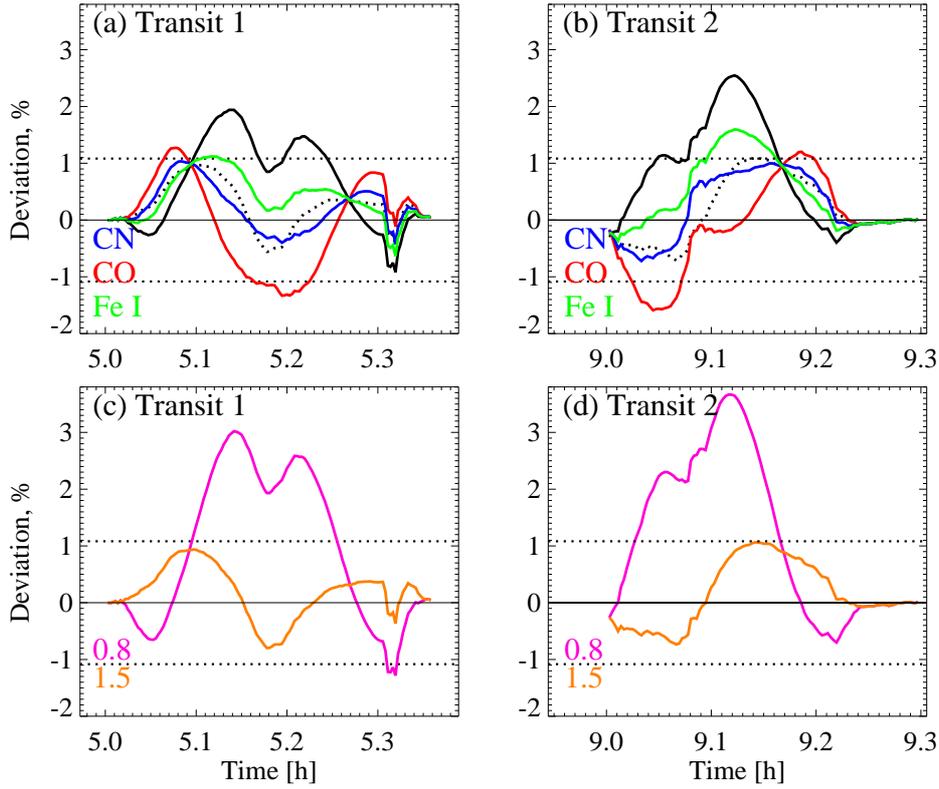}}
\caption{The same as Fig.~\ref{fig:ecl1_emp} but for different modes of calculations with the Model C (see caption to Fig.~\ref{fig:FALC}).  }\label{fig:FALC_pr}
\end{figure}

The above discussion is confirmed by the Figs.~\ref{fig:FALC} and \ref{fig:FALC_pr} where the CLV dependencies and the eclipse profiles for the different cases of the additional opacity scaling  are presented. One can see that the scaling of the additional opacity makes the CLV stronger. The scaling with the CO relative concentration results in too strong CLV and accordingly too small residual irradiance during the maximal phase of the eclipse. At the same time the scaling with the Fe I or CN relative concentrations moves the CLV dependencies and irradiance profiles very close to the observed ones  and hence can solve the problem addressed in the end of Sect.~\ref{subsect:T}. Let us notice that, in opposite to the calculations with the Model A, these calculations can also properly reproduce all main features in the solar spectrum.

 \begin{table}
 \caption{The same as Table~\ref{table:comp} but for different modes of calculations with the Model C and for the altered levels of the SOLSTCIE/SORCE measurements. The minimum values  of the discrepancy for different modes of calculations with the Model C are boldfaced.}\label{table:C}
 \begin{tabular}{c  c  c  c  c}     
\hline
     &  \multicolumn{2}{c}{Passage 1} &  \multicolumn{2}{c}{Passage 2} \\
     &  observed &   { empirical}  & observed &  { empirical}\\
 Standard   &        105.0       &   90.9    & 	129.3 	   &  111.5    \\
 CN              & 	   { 47.8}   &  	     { 19.1}        & 	 { 66.5}	   &    { \bf 25.0} \\
 CO 		  &         83.4      &   73.3	 &  85.4 	    & 	 60.8	         \\
 Fe I 		 &           59.5    &    	35.0  &	83.9   & 	  55.0	      \\
  80\% 		&   170.8     &   161.7    &    198.8  &    187.4\\  
  150\% 		&      51.4     &  15.6  &    	64.0 	   & 	  3.1	     \\
 \hline
\end{tabular}
\end{table}

While the relative variations of the irradiance can be measured by LYRA with a very high precision, the absolute calibration of LYRA radiometers is very tricky and renders the determination of  the  absolute level of the irradiance almost impossible without the use of an external reference. The additional opacity in the COSI were adjusted to reproduce the irradiance as measured by  SOLSTCIE/SORCE. If the solar UV irradiance were different from the measured by the SOLSTCIE/SORCE, then the different additional opacity would be necessary to reproduce it. The readjustment of the additional opacities will affect not only the absolute level of the measured irradiance but also the CLV.  Thus the CLV deduced from the eclipse analysis can be used to indirectly test the SOLSTCIE/SORCE measurements. With this goal we made two experiments readjusting the additional opacity so that COSI yields 80\% and 150\% of the SOLSTCIE/SORCE irradiance (the same scaling factor was applied for all wavelengths in the 160--320 nm interval). In these experiments we followed the approach of \opencite{shapiroetal2010} and assumed that the $f_c$ factor from the Eq.~(\ref{eq:factor}) is independent on height.

The corresponding CLV are given on Fig.~\ref{fig:FALC}, while the eclipse profiles are plotted on the bottom panels of Fig.~\ref{fig:FALC_pr}. One can see that the increase of the absolute level of the irradiance leads to a stronger CLV and good agreement with the measurements. Let us note that due to the problem of the standard modeling discussed in Sect.~\ref{subsect:T} this should be considered  not as a contradiction to the SOLSTICE/SORCE measurements but rather as a consequence of too weak CLV yielded by the calculations with the Model C and depth independent coefficient $f_c$. At the same time the decrease of the irradiance level leads to a significantly weaker CLV and strong deviations from the measurements.

In Table~\ref{table:C} we compare profiles which were discussed above with the LYRA measurements. The first line corresponds to the calculations with height independent additional opacity  adjusted to reproduce the SOLSTICE/SORCE measurements.

\section{Conclusions}\label{sect:conc}
The profiles of the eclipse light curves provide an important information for testing and refining the solar atmosphere models.
We have shown that the eclipse profiles observed by the Herzberg channel of LYRA are in a very good agreement with the synthetic profiles calculated with the 1D NLTE radiative transfer code COSI.

The calculated profiles are very sensitive to the temperature structure of the solar atmosphere and to the treatment of the UV opacity, whose significant part can still be missing from the modern models. The best agreement between the observed and measured profiles can be reached in two different regimes of calculations. The first one corresponds to the calculations  with the Model A atmosphere structure from \opencite{fontenlaetal1999} and constant coefficient of the additional opacity. The second one is the calculations with the Model C but  assuming that the missing opacity scales with the relative concentration of the neutral iron or CN molecule. Let us notice that while the calculations with the Model A are not able to properly reproduce the near UV, visible and infrared solar irradiance, the calculations with the model C yield a good agreement with measurements over the entire solar spectrum \cite{shapiroetal2010}. 

Our results could hint that the missing opacity originates in the layer a few hundred kilometers  below the temperature minimum and could be due to the unaccounted lines of the neutral iron or another element with the similar ionization potential (e.g. silicon or magnesium) or due to the unaccounted molecular lines (e.g. CN). 

We are aware of the limitations of modeling with 1D solar atmosphere. \linebreak \opencite{koesterkeetal2008} and \opencite{uitenbroek2011} show that 3D modeling can lead to a stronger CLV of the solar brightness.
This prevents us from making an unambiguous conclusion.     Nevertheless the 1D radiative transfer code are widely used for the interpretation of the the stellar and solar spectra and  presently they  are the  {\it de facto} standards for modeling the solar spectral irradiance variability
\cite{krivovasolanki2008,fontenlaetal2009,domingo2009}.

 { The variability of the solar irradiance on the 11-year and solar rotational time-scales is usually attributed to the competition between the irradiance increase due to the bright active components (e.g. plage and active network) and irradiance  decrease due to the dark sunspots (see e.g. \opencite{krivovaetal2003}). The contrast between active features and the quiet Sun strongly depends on the disk position \cite{unruhetal1999, fliggeetal2000}. Thus the center-to-limb variations of the solar brightness analyzed in this paper  play an important role in the modeling of the solar irradiance variability.   The fact that the calculations with COSI are in a good agreement with the measurements strongly supports its suitability for such modeling \cite{anna_eclipse}.}

\begin{acks}
The research leading to this paper was supported by the Swiss National Science Foundation under grant CRSI122-130642 (FUPSOL) and grant 200020-130102. We thank the  LYRA PROBA 2 science team for their work in producing the data sets used in this paper and their helpful recommendations.
\end{acks}

\bibliographystyle{spr-mp-sola}

\end{article} 
\end{document}